\documentclass[aps,prl,reprint,superscriptaddress]{revtex4-2}
\usepackage{graphicx}  
\usepackage{subcaption} 
\usepackage{amsmath}   
\usepackage{amssymb}   
\usepackage{hyperref}  
\usepackage{bm}        
\usepackage{braket}
\usepackage{float}
\graphicspath{{./figures/}}

\begin{document}

\title{Quantum Beatings in Optical Cavities}

\author{Ishaan Ganti}
\email{ishaan\_ganti@brown.edu}
\affiliation{Brown University, Providence, Rhode Island, 02912, USA}
\affiliation{Department of Chemistry, Massachusetts Institute of Technology, Massachusetts, 02139, USA}
\author{Jianshu Cao}
\email{jianshu@mit.edu}
\affiliation{Department of Chemistry, Massachusetts Institute of Technology, Massachusetts, 02139, USA}

\date{\today}

\begin{abstract}
    Cavity polaritons, quasiparticles formed by coherent light-matter coupling, 
    are at the heart of fundamental concepts of quantum optics. The quintessential signature of this coherent coupling is the Rabi oscillation, 
    which results from the neglect of the counter-rotating-wave (CRW) effect
    in the weak-coupling regime.  The goal of this letter is to predict resonant beatings that envelop the Rabi oscillation on the second or higher excitation manifold.   These polariton beatings arise from the CRW term in the Dicke or Pauli-Fierz model 
    and are directly correlated with the asymmetry in polariton eigenenergies. 
    Our findings highlight the relevance of the CRW effect even in the weak-coupling regime, offer novel perspectives about coherent polariton dynamics, and shed new light on experiments of coupled quantum systems.
\end{abstract}

\maketitle

\section{Introduction}
Cavity polaritons, quasiparticles arising from the coupling of 
molecules and photons in an optical cavity, have become a key concept in quantum optics.\cite{hopfield58,forn19}
Along with superconducting circuits, trapped ions, and cold atoms in optical lattices, these hybrid quantum platforms
have lent themselves to a diverse set of applications in quantum information,
materials science, and condensed matter physics. 
Central to their theoretical descriptions are fundamental models describing the interaction 
between an ensemble of two-level systems (TLS) 
and the cavity photons, namely, the Tavis-Cummings (TC) model, the Dicke model (DM), 
and the Pauli-Fierz (PF) model. Common to these models is the prediction of Rabi oscillations, 
which are widely recognized
as the quintessential signature of quantum coherence in the light-matter interaction.\cite{ebbesen16,garcia21}
In this letter, we predict a more subtle quantum signature that goes beyond the Rabi oscillation
in revealing the fine energy structure of light-matter hybrid states (i.e., polaritons) and
in differentiating between various theoretical descriptions, even in the weak coupling regime.

\section{Model Hamiltonians}

Using dimensionless units, the symmetrized TC model Hamiltonian\cite{tavis68} is given by
\begin{gather}
        H_{TC} = \omega_m J_z + 
    \omega_c a^{\dag }a + 
    \frac{g}{\sqrt{N}} \left( J_{+}a + J_{-}a^{\dag } \right) 
    \label{TC}
\end{gather}
where $\omega_m$ and $\omega_c$ are the TLS and cavity frequencies, 
$N$ is the number of TLS, $g$ is the
TLS-cavity coupling strength, $a$ and $a^{\dag }$ are the creation
and annihilation operators of the cavity mode, and
$J_{\pm }$ are the collective raising and lowering operators for the symmetrized TLS.
Notably, the interaction term in the TC Hamiltonian ensures the conservation of 
excitation number of the polariton system.
 Next, the Dicke Hamiltonian \cite{dicke54} is given by 
\begin{gather}
    H_{DM} = \omega_m J_z + 
    \omega_c a^{\dag }a + 
    \frac{g}{\sqrt{N}} J_{x}  \left( a + a^{\dag } \right) 
    \label{DM}
\end{gather}
where $J_x = J_{+} + J_{-}$. Eq.~(\ref{DM}) introduces the counter-rotating-wave (CRW) term, i.e.
$H_{CRW} = g(J_{+}a^{\dagger} + J_{-}a) / \sqrt{N}$, which breaks the conservation of excitations. 
Finally, the PF Hamiltonian \cite{power59} is given by 
\begin{gather}
    H_{PF} = H_{DM} + \frac{g^2}{\omega_c N} \left( J_x \right)^2
    \label{PF}
\end{gather}
where the additional term is the 
dipole self-energy (DSE) accounting for cavity-mediated interactions between
the TLS. In the weak-coupling regime, the RWA is commonly used to simplify the analysis
by neglecting the CRW term. The conditions for the RWA are 
$\omega_c - \omega_m = \Delta \omega \ll \omega_c + \omega_m$ 
(resonance or near-resonance) and $g \ll \omega_m$.

\section{Numerical Analysis}
Numerical analysis of the cavity system involves the photon counts, spectral decomposition and energy schema. 
Simulations were carried out for $N$ from 1 to 15, but we focus primarily 
on the $N=2$ resonant case for the convenience of the subsequent perturbative analysis. The parameters
are $\omega_c = \omega_m = 1.0$, $0.01 \le g \le 0.12$, and
an initial state consisting of a TLS ensemble in the doubly-excited symmetric state and the
cavity photon in the vacuum state, $\psi(0)= \ket{s_2, 0}$. To propagate the polariton system, we combine
numerical integration of the Schrödinger equation and spectral decomposition [see Eq.~(\ref{decomposition})].

\begin{figure*}[!htbp]
    \centering
    \begin{subfigure}[b]{0.32\textwidth}
        \includegraphics[width=\textwidth]{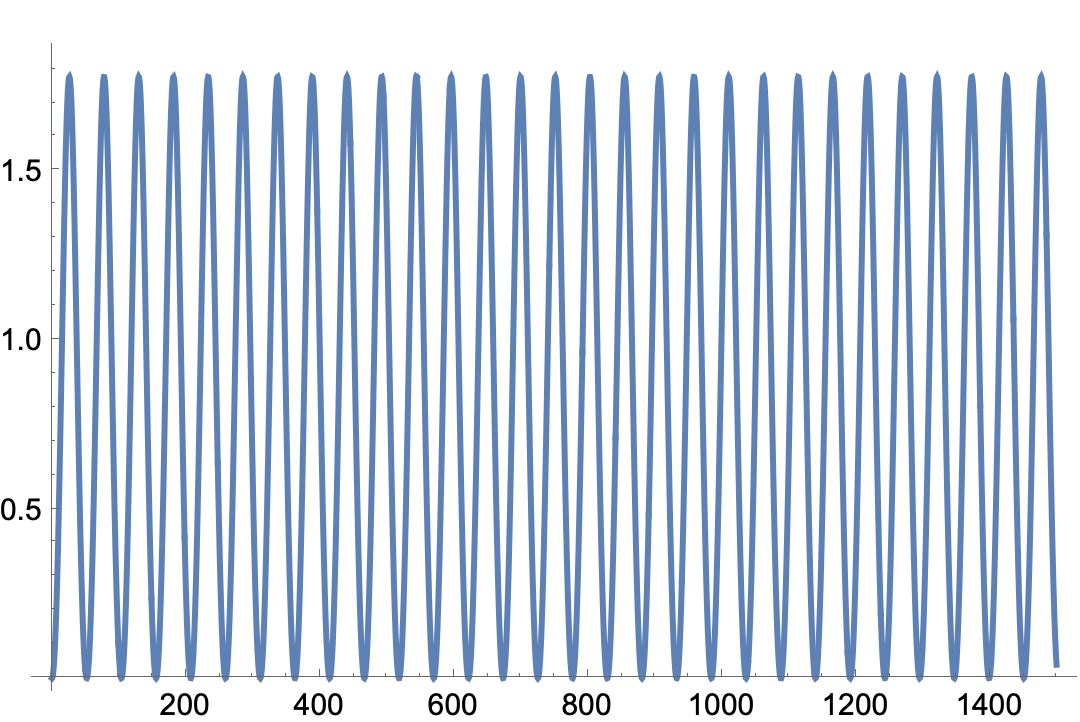}
        \caption{TC Model}
        \label{fig:tc_model}
    \end{subfigure}
    \hfill
    \begin{subfigure}[b]{0.32\textwidth}
        \includegraphics[width=\textwidth]{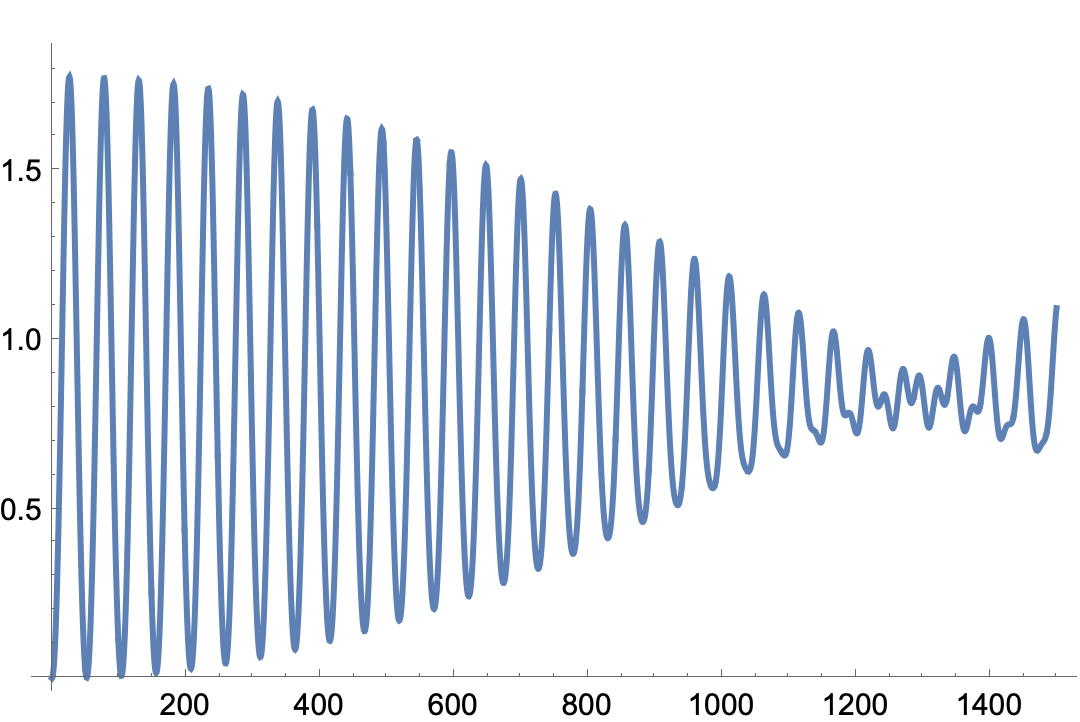}
        \caption{Dicke Model}
        \label{fig:dicke_model}
    \end{subfigure}
    \hfill
    \begin{subfigure}[b]{0.32\textwidth}
        \includegraphics[width=\textwidth]{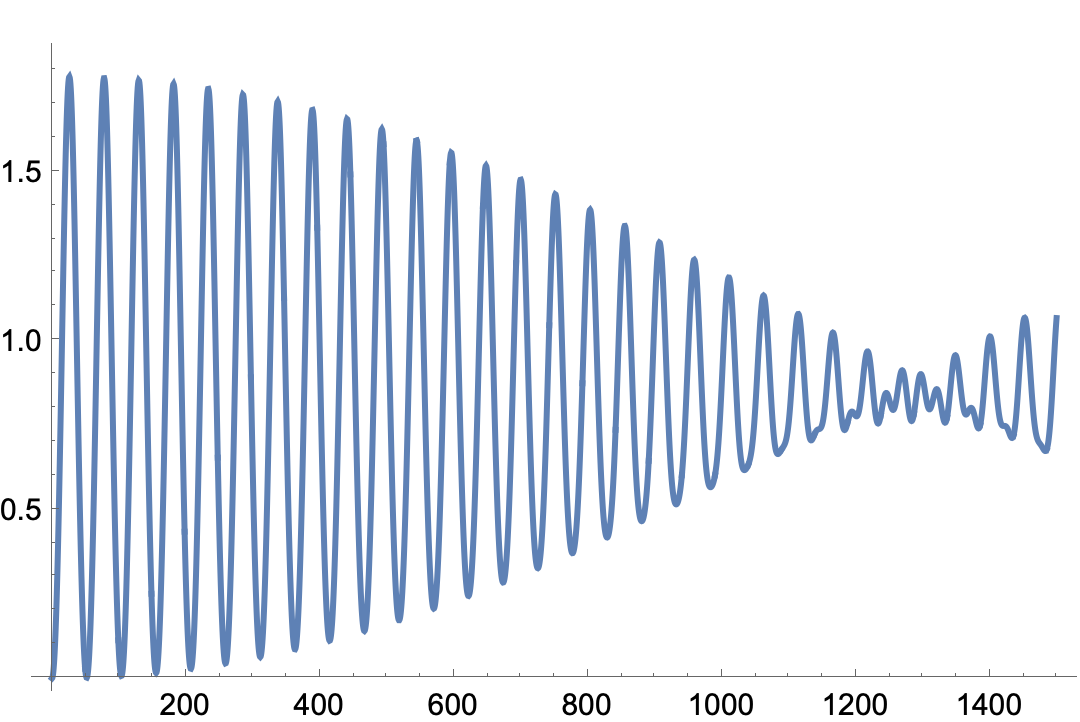} 
        \caption{PF Model}
        \label{fig:pf_model}
    \end{subfigure}
    \caption{Average photon counts for 
    all three models. Parameters: $g = 0.07$, $\omega_m = \omega_c = 1$, 
    $N = 2$.} 
    \label{fig:photon_expectation_time}
\end{figure*}

\subsection{Average Photon Count}
The average photon count, i.e., the expectation value of photon number operator $\left < a^{\dag }a \right >$,  is plotted
for the TC, Dicke, and PF models in Fig.~\ref {fig:photon_expectation_time}. These plots illustrate
the presence of polariton beatings in the Dicke and PF models but not in the resonant TC model. 
Additionally, the beating dynamics in the Dicke and PF models are
similar--the only noticeable difference between the two occurs between 
$t=1200$ and $t=1400$, where the oscillation
amplitude diminishes. The similarity can be attributed to the
DSE term in the PF model, which scales with $g^2 / N$, deeming it an 
order of magnitude less significant than the CRW terms. Thus, we
will focus on the Dicke model in the subsequent study of the beating dynamics. 
As shown in SM, a similar beating pattern 
can be observed on the higher excitation manifold.  Further, 
the variance of the photon count also exhibits a complex beating pattern as shown in SM,
suggesting quantum features in photon statistics.\cite{cao205,cao219}

\begin{figure*}[!tp]
    \centering
    \begin{subfigure}[b]{0.32\textwidth}
        \centering
        \includegraphics[width=\textwidth]{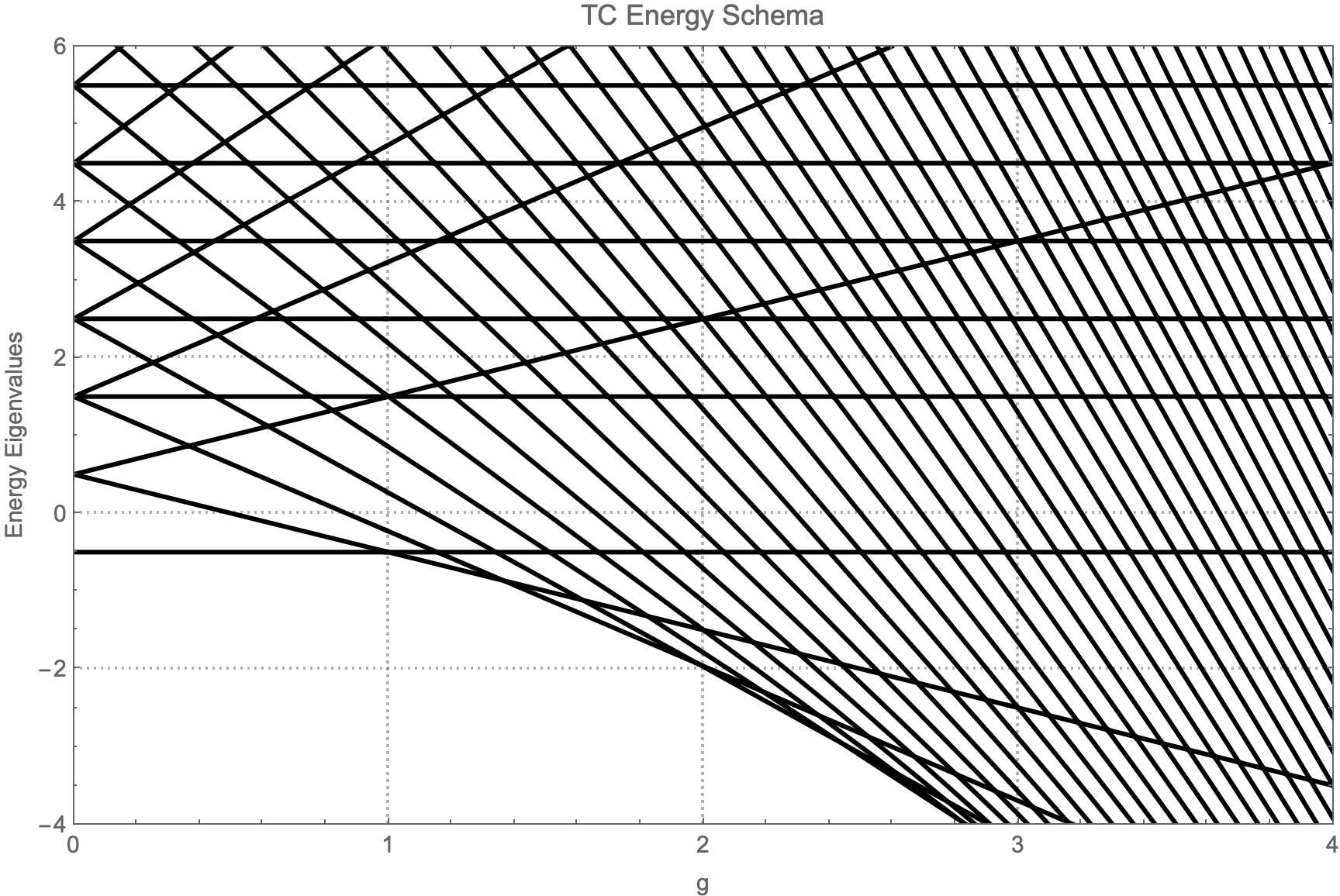}
        \label{fig:tc_model_schema}
    \end{subfigure}
    \hfill
    \begin{subfigure}[b]{0.32\textwidth}
        \centering
        \includegraphics[width=\textwidth]{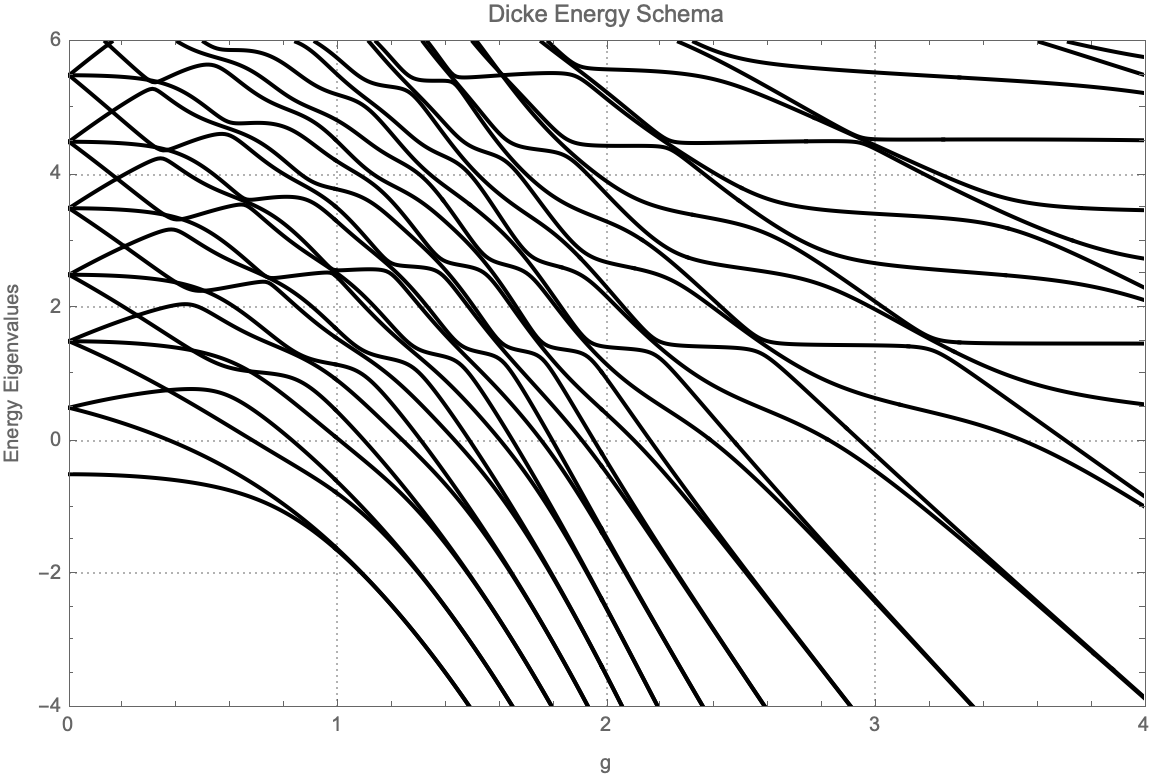}
        \label{fig:dicke_model_schema}
    \end{subfigure}
    \hfill
    \begin{subfigure}[b]{0.32\textwidth}
        \centering
        \includegraphics[width=\textwidth]{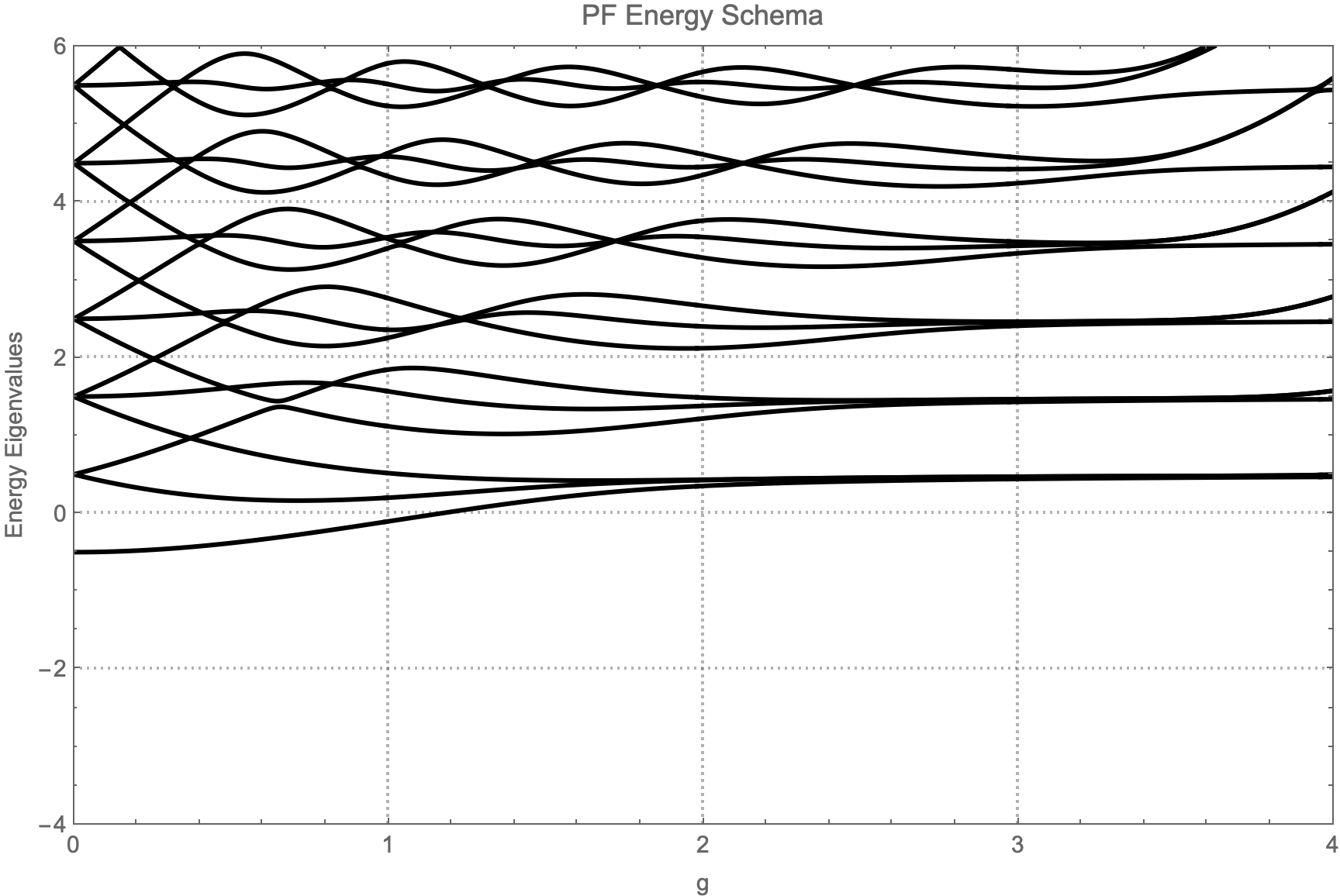}
        \label{fig:pf_model_schema}
    \end{subfigure}
    \caption{The energy schema for the TC, Dicke, and PF models
    for $N=2$. The Hamiltonians are both scaled on-resonance with 
    $g = 0.07$. Note the agreement between the three schema
    for small $g$.} 
    \label{fig:energy_schema}
\end{figure*}

\subsection{Spectral decomposition}

The time-evolution of the polariton wave function can be evaluated via spectral decomposition,
\begin{equation}
\psi(t)= \sum_{\lambda} \mid P_{\lambda} \rangle  e^{-i E_{\lambda} t}  \langle P_{\lambda} \mid  \psi(0) \rangle =
\sum_{\lambda}  c_{\lambda} e^{-i E_{\lambda} t}  \mid P_{\lambda} \rangle 
\label{decomposition}
\end{equation}
where $\psi(0)$ is the initial state and $c_{\lambda}=  \langle P_{\lambda} \mid  \psi(0) \rangle$ is the projection coefficient. 
In Eq.~(\ref{decomposition}), 
$\lambda$ denotes a polariton eigen-state with the eigen-energy $E_{\lambda}$ and eigen-vector $P_{\lambda}$.

As suggested by Eq.~(\ref{decomposition}), 
the beating can be caused by the eigenenergy $E_{\lambda}$ and eigenvector $P_{\lambda}$ as well as by the projection coefficient $c_{\lambda}$.  
Interestingly, reconstructing a wavefunction using TC eigenvalues with Dicke eigenvectors and Dicke projection coefficients yields virtually no beatings. Instead,  using Dicke eigenvalues with TC eigenvectors and TC projection coefficients accurately reproduces the observed beating in the standard Dicke model (see Fig.~1 in Supplementary Material, i.e., SM). Thus, the quantum beatings in the Dicke model are caused by the energy shifts due to the CRW term and are well approximated by 
\begin{equation}
\psi(t)  =
\sum_{\lambda}  c_{\lambda} e^{-i E_{\lambda} t}  \mid P_{\lambda} \rangle
\approx 
\sum_{\lambda}  c^{TC}_{\lambda} e^{-i E^{DM}_{\lambda} t}  \mid P^{TC}_{\lambda} \rangle
\label{approx}
\end{equation}
where $c^{TC}$ and $P^{TC}$ are based on the TC model
and $E^{DM}$ are the DM eigenenergies.

\subsection{Energy Schema \& Eigenvalue Asymmetry}

The energy schema for the TC, Dicke, and PF models in Fig. ~\ref{fig:energy_schema}
show the agreement between the three models for small $g$ and increasing energy 
shifts for large $g$. Note the symmetric structure in the TC model, the eigen-energy coalescence in the Dicke model, 
and the lower bound in the PF model.\cite{dicke54,jaynes63,gross82,shore93,emary03,schafer20a,taylor20,cao221}
A less noticed feature is the eigenvalue asymmetry,  defined as
the difference between the middle polariton energy, $E_0$, and the average of the outer polariton energies, $(E_++E_-)/2$, i.e., 
\begin{equation}
\alpha = { E_+ + E_- \over 2} -E_0. 
\end{equation}
Evidently, the energy schema of the TC model is symmetric by construction, 
so $\alpha=0$ under the RWA in the TC model, but it grows with $g$ for the Dicke model, 
suggesting that the asymmetry $\alpha$ drives the quantum beating.

\section{Photon count in the TC model}

Within the second excitation manifold (SEM), we construct the symmetrized bright basis
\begin{gather}
      M_2^s= \{ \ket{s_0, 2}, \, \ket{s_1, 1}, \,  \ket{s_2, 0} \}
\end{gather}
where the first letter denotes the symmetric molecular state, 
 with the subscript specifying the excitation level,
 and the second digit denotes the photon quantum number. Then, the SEM Hamiltonian becomes
\begin{gather}
        \begin{pmatrix} 2 \omega_c & \Omega_{N} & 0 \\
        \Omega_{N} & \omega_c + \omega_m & \Omega_{N-1} \\
    0 & \Omega_{N-1} & 2\omega_m \end{pmatrix} 
\end{gather}
where $\Omega_M = g \sqrt{2M / N}$.
Solving this Hamiltonian in general is possible but tedious; we opt to
consider the resonant case, $\omega_m=\omega_c$. \cite{cao221}
Then, we obtain three
polariton solutions, $\{ P_+, P_-, P_0 \}$, with corresponding eigenvalues
\begin{gather}
        E_0 = 2\omega , \quad 
    E_{\pm} = 2\omega \pm \Omega
    \label{eq:tc_polariton_sols}
\end{gather}
where $\Omega= \sqrt{\Omega_N^2 + \Omega_{N-1}^2}= \Omega_{2N-1}$ is the effective Rabi frequency of the SEM.
Notably, the upper and lower polariton energies average to exactly the middle
polariton energy such that $\alpha=0$.  
Next, we evaluate  the time-dependence of the cavity photon expectation value and obtain
\begin{gather}
 \left < n \right > (t) =   -\frac{2(N-1)(1-4N + \cos (\Omega \, t)) 
    }{(2N-1)^2}
    \sin^2 \left(\frac{\Omega \,  t}{2}\right)
    \label{eq:tc_photon_num}
\end{gather}
which is derived in the SM. 
Eq.~(\ref{eq:tc_photon_num}) predicts the Rabi oscillations in
the TC model accurately.

\begin{figure*}[!htbp]
    \centering
    \begin{subfigure}[b]{0.46\textwidth}
        \centering
        \includegraphics[width=\textwidth]{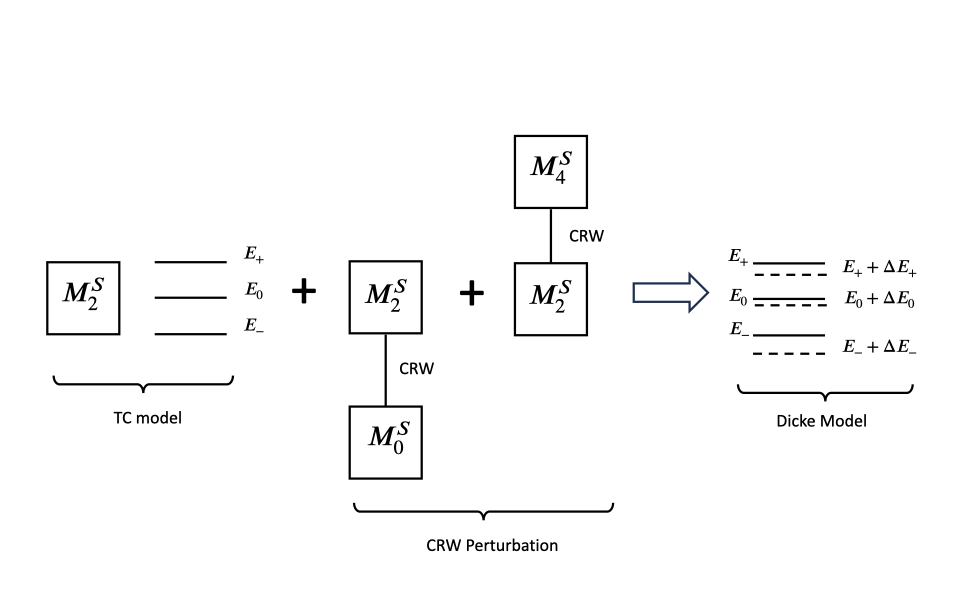}
        \caption{An outline of the perturbative approach. $M_n^s$ denotes the symmetric polariton states on the n-th
        excitation manifold.}
        \label{fig:pertubative_approach}
    \end{subfigure}
    \begin{subfigure}[b]{0.46\textwidth}
        \centering
        \includegraphics[width=\textwidth, trim=0 0 0 130, clip]{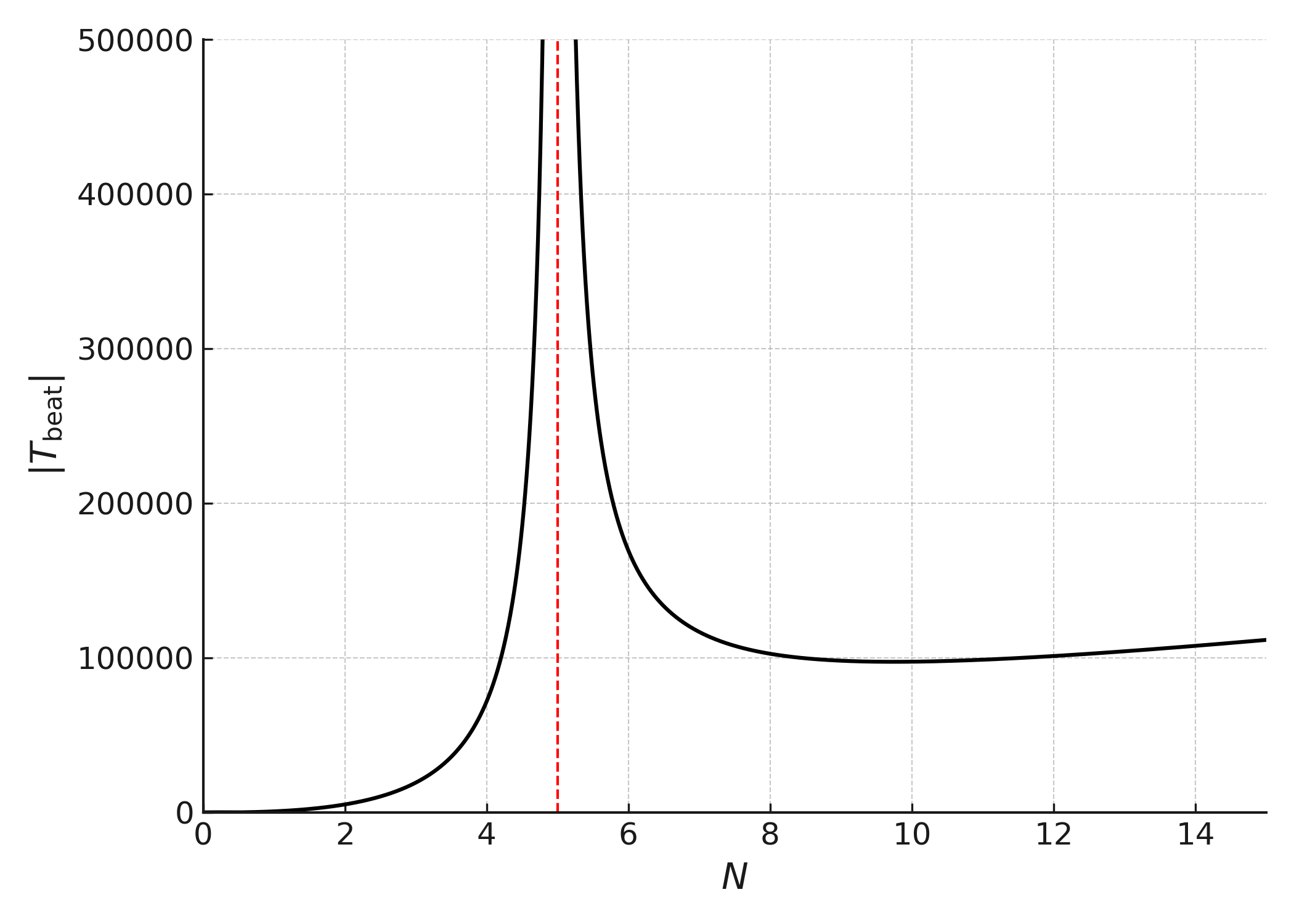}
        \caption{Beating period as a function of N, as predicted by Eq.~\ref{eq:beating-period}. A local minimum is predicted between $N=9$ and $N=10$, and a divergence is predicted at N=5.}
        \label{fig:T-beat}
    \end{subfigure}
    \caption{}
\end{figure*}

\section{Perturbative analysis of beating}

We proceed by deriving the beating dynamics and verifying the energy asymmetry hypothesis. 
To approximate an analytical form for the eigen-energy of the Dicke model, 
we perturb the TC eigen-energies by the CRW term.\cite{cao221}
First, we consider the coupling between the global ground state and the three polariton states $\{P_+, P_-, P_0 \}$ on the SEM. 
The middle polariton $P_0$ is orthogonal to
$\ket{s_1, 1}$, so it is not affected. The upper and lower polaritons, $P_{\pm}$, are perturbed as
\begin{gather}
    \Delta E_{\pm}^{(0)} = 
    \frac{\left | \braket{P_{\pm}|H_{CRW}|s_0, 0} \right |^2}{E_{\pm}}
    = 
    \frac{g^2}{2(2\omega \pm \Omega)}
    \label{eq:ground_upper_polariton}
\end{gather}
which leads to accurate beating dynamics for N=2 (see SM).

Next, we consider the coupling to the fourth-excitation manifold (FEM) due to the CRW term $J_{+}a^{\dag}$ .  To circumvent diagonalizing the TC Hamiltonian on the FEM,  we adopt the product states
instead of the hybrid states, which will shift the unperturbed energy $4\omega$ by the order of $\Omega$. 
Since $\Omega \propto g$, within the weak coupling regime, $\Omega / \omega$  is small, so we approximate all FEM eigenenergies by $4\omega$.   Then, we write the perturbation on the SEM eigenenergies due to the coupling to the FEM as 
\begin{gather}
    \Delta E_i^{(4)} = \sum_{n=1}^{3}
    \frac{|\braket{s_n, 4-n|J_{+}a^{\dag}|P_i}|^2}{E_i - 4\omega} 
    \label{eq:FEM_sum}
\end{gather}
where the resonance condition $E_{s_1, 3} = E_{s_2, 2} = 
E_{s_3, 1} = 4\omega$ is used.  As detailed in the SM, 
the FEM-perturbed energies are
\begin{gather}
    \Delta E_{0}^{(4)} = \frac{3g^2(3 - 2N)}{2\omega(2N-1)}
    \label{eq:eigenenergy_FEM_shifts_1}\\
    \Delta E_{\pm }^{(4)} = 
    \frac{g^2(10 + 7N(2N-3))}{2N(2N-1)(- 2\omega)}
    \label{eq:eigenenergy_FEM_shifts_2}
\end{gather}
We then define $\Delta E_i = \Delta E_i^{(0)} + \Delta E_i^{(4)}$
such that the perturbed polariton eigenenergies in the Dicke model are
\begin{gather}
    E_0 = 2\omega + \Delta E_0, \quad 
    E_{\pm} = 2\omega \pm \Omega
    + \Delta E_{\pm}
    \label{eq:eigenenergy_FINAL}
\end{gather}

Motivated by Eq.~(\ref{approx}), we opt to keep the unperturbed
TC eigenstates and projection coefficients, but adopt
the perturbed DM eigenenergies. We  make a further approximation by noticing that
$|\Delta E_{-}^{(0)} - \Delta E_{+}^{(0)}|$ is generally small,  so $\Delta E_{-}^{(0)} \approx \Delta E_{+}^{(0)} $.
With these approximations, the average photon number is given by  
\begin{gather}
\left\langle n \right\rangle (t) = 
\frac{N - 1}{2 (2N - 1)^2} \Bigg[ 
    \cos(2\, \Omega\, t) + 8N - 1 \nonumber \\
    \qquad -\, 8N\, \cos(\alpha t)\, \cos(\Omega\, t) 
\Bigg]
\label{eq:dicke_photon_num_approx}
\end{gather}
which reproduces the numerically exact results in the relevant parameter range. 
The above analysis allows us to identify the beating frequency as
\begin{equation}
\alpha  \approx  { g^2 \over 2 \omega} { N-5 \over N(2N-1) }
\label{alpha}
\end{equation}
which is a key result of this study.

\begin{figure*}[!htbp]
    \centering
    \begin{subfigure}[b]{0.32\textwidth}
        \includegraphics[width=\textwidth]{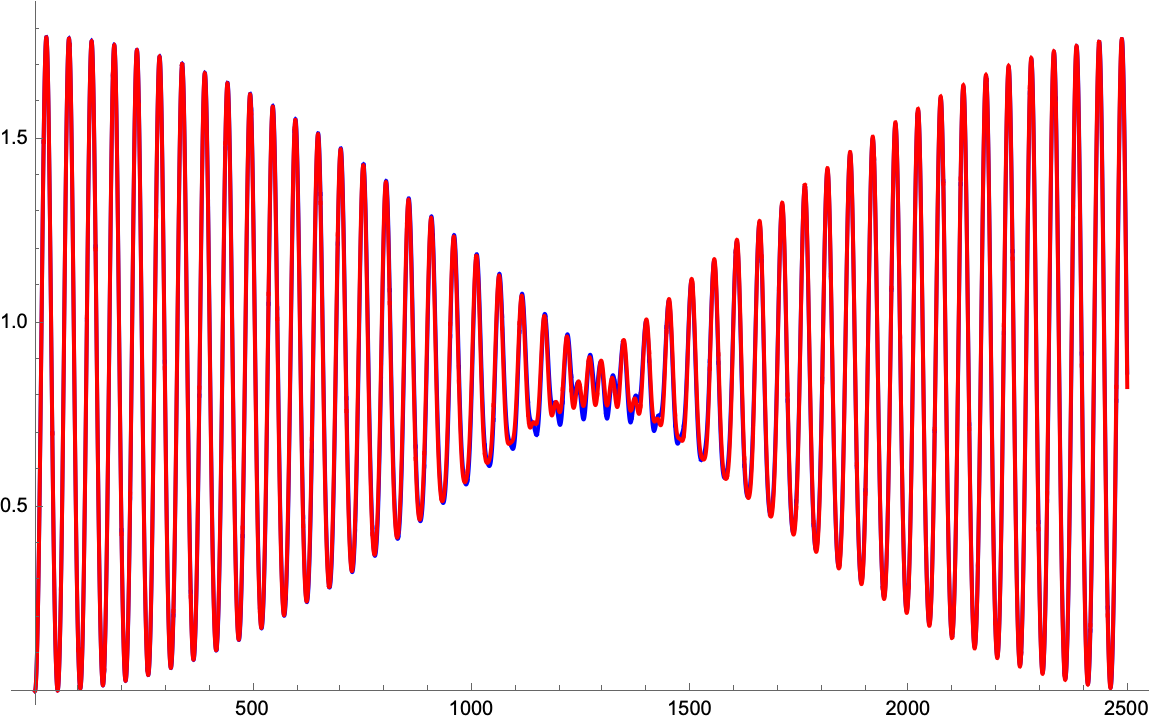}
        \caption{$N = 2$}
        \label{fig:dicke_approx_GOOD_N=2}
    \end{subfigure}
    \hfill
    \begin{subfigure}[b]{0.32\textwidth}
        \includegraphics[width=\textwidth]{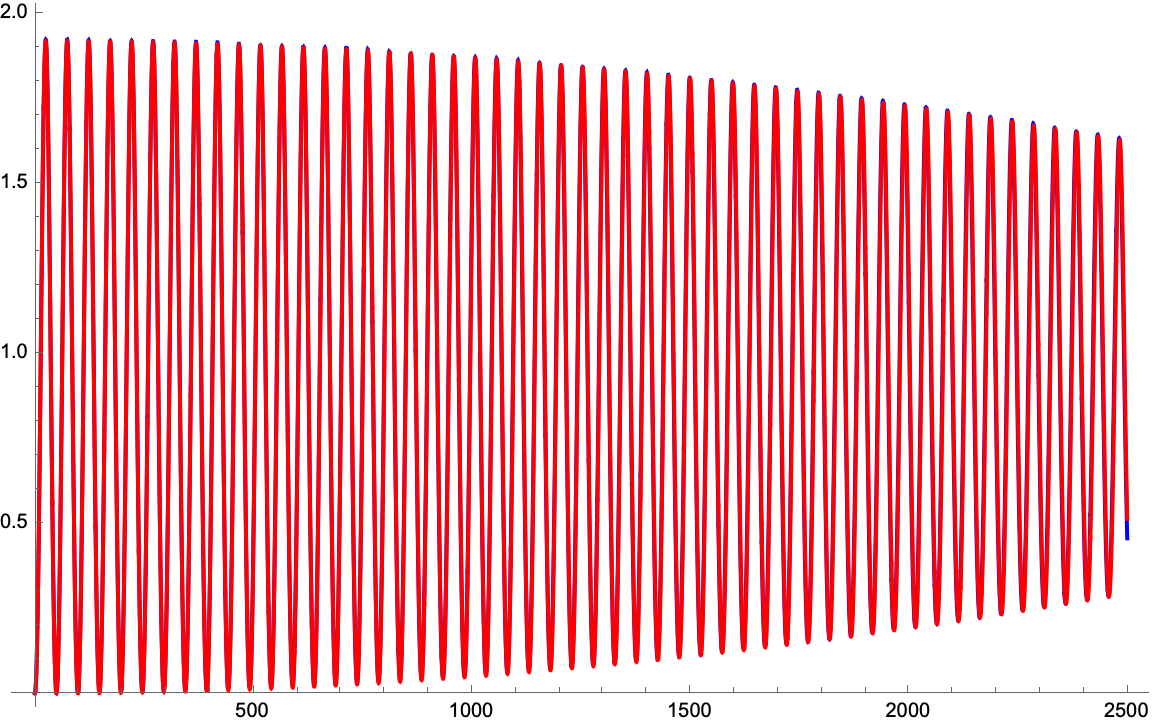}
        \caption{$N = 3$}
        \label{fig:dicke_approx_GOOD_N=2}
    \end{subfigure}
    \hfill
    \begin{subfigure}[b]{0.32\textwidth}
        \includegraphics[width=\textwidth]{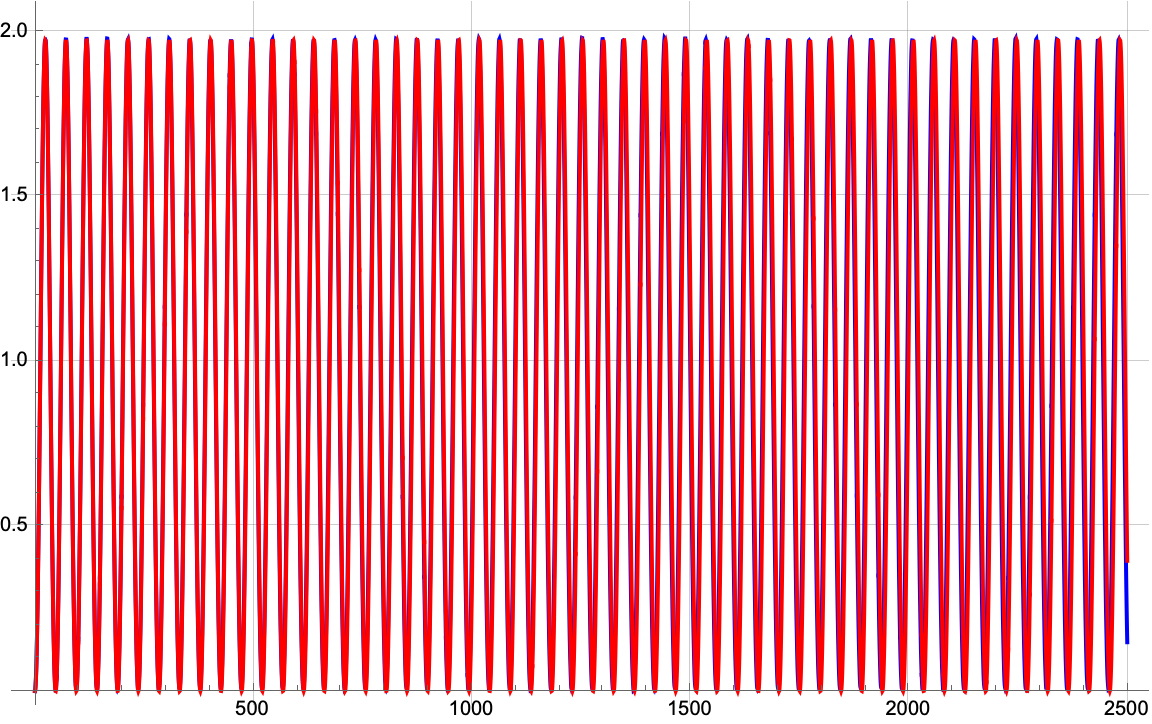}
        \caption{$N = 5$}
        \label{fig:dicke_approx_GOOD_N=5}
    \end{subfigure}
    \caption{Average photon count for the Dicke model: 
    numerical results (blue) against the analytical approximations in 
    Eq. ~\ref{eq:dicke_photon_num_approx} (red).
    Parameters: $g = 0.07$, $\omega_m = \omega_c = 1$. Note the lack of beating in plot (c).}
    \label{fig:dicke_approx_combined}
\end{figure*}

Eq.~(\ref{alpha}) predicts the g and N dependence of the polariton beating.
For example, the beating frequency scales quadratically with g, the Rabi frequency 
scales linearly, and thus their ratio $\alpha/\Omega$ is linear in g. 
As shown in Fig.~3 of the SM, the polariton beating becomes more accessible 
as g increases. Next, following Eq.~(\ref{alpha}), we obtain the beating period as
\begin{gather}
    T_{\text{beat}} = { 2\pi \over \alpha} \approx \frac{4\pi \omega}{g^2} \left |
    \frac{N(2N-1)}{N-5}
    \right | \label{eq:beating-period}
\end{gather}
which predicts the maximum beating for
$N=2$ and, interestingly, no beating for $N=5$.  Numerical confirmation of this prediction is shown in Fig.~\ref{fig:dicke_approx_combined}, and the beating frequency
is plotted as a function of $N$ in Fig.~\ref{fig:T-beat}.

\section{Experimental Relevance}
Notably, modern experimental setups are capable of realizing the prediction of polariton beating. Consider a feasible physical system \cite{majer07} with parameters $\frac{\omega_c}{2\pi} \approx 6$
GHz, $\frac{g}{2\pi} \approx 450$ MHz, which gives $\frac{g}{\omega_c} \approx 0.075$, precisely within the coupling range of our calculations. Further, we consider just two TLSs in the cavity; the Rabi oscillation
period is then $T_{Rabi} \approx 1.56 \text{ ns}$.  Per Fig. ~\ref{fig:photon_expectation_time}, it
takes roughly 25 coherent Rabi oscillations for a fully
excited initial state to be maximally enveloped by the
beating frequency. Experimentally realizing this
would require a polariton coherence time approximately
given by 
\begin{equation}
    { T_{\text{beat}}  \over T_{\text{Rabi}} } \approx
    25 \implies 
    T_{\text{coherent}} = T_{\text{beat}} \approx 39.3 \text{ ns}
\end{equation}
The qubit decay rates and cavity decay rates of
modern circuit QED systems allow for coherence times comfortably greater than 100 ns \cite{place21}, demonstrating the relevance of our predictions. Experimental design of optical cavities, including those under vibrational strong coupling,\cite{george16,xiang18},  
can also approach this regime. 

\section{Discussions}

To explore the validity and implications of our prediction, we discuss the following issues:
(1) off-resonant effects; (2) static disorder and dissipation; (3) quantum-classical correspondence.

\begin{enumerate}
\item 

The analysis presented in the letter is for the resonance case.
Under the near-resonance condition, i.e. 
$\omega_c \neq \omega_m$ but $\omega_c - \omega_m \approx 0$, 
the TC model also causes noticeable beatings. 
Thus, the observed beating combines the contributions from the CRW term and detuning.
As shown in the SM , we can differentiate their effects by their distinct sign dependence:
The detuning contribution depends on the sign of detuning, whereas the CRW contribution is sign-independent.
In fact, we can adjust the detuning to cancel the CRW contribution
such that the beating vanishes. 

\item
The current calculation is based on the Hamiltonian description of isolated
cavity polaritons and can be extended to incorporate various dissipative channels, including static disorder,\cite{cao204,cao211} 
stochastic noise (i.e. phonon couplings),\cite{cao223} and photon loss.\cite{cao217}  For example, 

Further, a recent calculation of disordered polariton dynamics shows negligible effects 
on SEM polariton dynamics until the strength of static disorder reaches a critical value.\cite{cao217} 
Therefore, it is reasonable to speculate that static disorder is perturbative below the critical value and various dissipative channels can be treated additively. 

\item 
Finally, we discuss the implications of polariton beatings in the context of quantum-classical correspondence.
As the textbook signature of light-matter coherence, 
Rabi oscillations or splittings have been studied within the semiclassical (e.g., Floquet)
or mixed quantum-classical (MQC) framework.
The polariton beating predicted in this paper is a subtle signature of 
light-matter coherence, and thus provides a stringent test of these approximations. 
The issue is worth further investigation given the extensive use of the MQC method in 
simulating complex molecular systems in optical cavities.\cite{hoffmann19b,li19,hsieh25} 
Along this line, an analogy can be established between polariton beatings
analyzed here and anharmonic vibrations studied before.\cite{cao54,cao86,gruenbaum08,dutta25}
It stands to reason that Rabi oscillations are analogous to harmonic response whereas polariton beatings
are analogous to anharmonic response, which requires more advanced treatments of quantum-classical correspondence.

\end{enumerate}

\section{Conclusion}

This letter predicts the emergence of quantum beatings in cavity polaritons.
The characteristic oscillatory behavior arises from the CRW term  
in the Dicke or PF model and is thus absent in the resonant TC model. 
Our analysis demonstrates that the asymmetric shifts in the eigenenergies drive these beatings within the initial excitation manifold. The polariton beatings are fundamentally different from
the Rabi oscillations, as they appear on longer time scales, vanish within the single-excitation manifold, 
and define a unique signature of the CRW effect.  Experimental verification can be realized in optical lattices, superconducting circuits, 
and optical cavities including these under vibrational strong coupling.

\bibliography{polariton, pub_cao}

\begin{thebibliography}{31}%
\makeatletter
\providecommand \@ifxundefined [1]{%
 \@ifx{#1\undefined}
}%
\providecommand \@ifnum [1]{%
 \ifnum #1\expandafter \@firstoftwo
 \else \expandafter \@secondoftwo
 \fi
}%
\providecommand \@ifx [1]{%
 \ifx #1\expandafter \@firstoftwo
 \else \expandafter \@secondoftwo
 \fi
}%
\providecommand \natexlab [1]{#1}%
\providecommand \enquote  [1]{``#1''}%
\providecommand \bibnamefont  [1]{#1}%
\providecommand \bibfnamefont [1]{#1}%
\providecommand \citenamefont [1]{#1}%
\providecommand \href@noop [0]{\@secondoftwo}%
\providecommand \href [0]{\begingroup \@sanitize@url \@href}%
\providecommand \@href[1]{\@@startlink{#1}\@@href}%
\providecommand \@@href[1]{\endgroup#1\@@endlink}%
\providecommand \@sanitize@url [0]{\catcode `\\12\catcode `\$12\catcode
  `\&12\catcode `\#12\catcode `\^12\catcode `\_12\catcode `\%12\relax}%
\providecommand \@@startlink[1]{}%
\providecommand \@@endlink[0]{}%
\providecommand \url  [0]{\begingroup\@sanitize@url \@url }%
\providecommand \@url [1]{\endgroup\@href {#1}{\urlprefix }}%
\providecommand \urlprefix  [0]{URL }%
\providecommand \Eprint [0]{\href }%
\providecommand \doibase [0]{https://doi.org/}%
\providecommand \selectlanguage [0]{\@gobble}%
\providecommand \bibinfo  [0]{\@secondoftwo}%
\providecommand \bibfield  [0]{\@secondoftwo}%
\providecommand \translation [1]{[#1]}%
\providecommand \BibitemOpen [0]{}%
\providecommand \bibitemStop [0]{}%
\providecommand \bibitemNoStop [0]{.\EOS\space}%
\providecommand \EOS [0]{\spacefactor3000\relax}%
\providecommand \BibitemShut  [1]{\csname bibitem#1\endcsname}%
\let\auto@bib@innerbib\@empty
\bibitem [{\citenamefont {Hopfield}(1958)}]{hopfield58}%
  \BibitemOpen
  \bibfield  {author} {\bibinfo {author} {\bibfnamefont {J.~J.}\ \bibnamefont
  {Hopfield}},\ }\bibfield  {title} {\bibinfo {title} {Theory of the
  contribution of excitons to the complex dielectric constant of crystals},\
  }\href@noop {} {\bibfield  {journal} {\bibinfo  {journal} {Physical Review}\
  }\textbf {\bibinfo {volume} {112}},\ \bibinfo {pages} {1555} (\bibinfo {year}
  {1958})}\BibitemShut {NoStop}%
\bibitem [{\citenamefont {Forn-D{\'\i}az}\ \emph {et~al.}(2019)\citenamefont
  {Forn-D{\'\i}az}, \citenamefont {Lamata}, \citenamefont {Rico}, \citenamefont
  {Kono},\ and\ \citenamefont {Solano}}]{forn19}%
  \BibitemOpen
  \bibfield  {author} {\bibinfo {author} {\bibfnamefont {P.}~\bibnamefont
  {Forn-D{\'\i}az}}, \bibinfo {author} {\bibfnamefont {L.}~\bibnamefont
  {Lamata}}, \bibinfo {author} {\bibfnamefont {E.}~\bibnamefont {Rico}},
  \bibinfo {author} {\bibfnamefont {J.}~\bibnamefont {Kono}},\ and\ \bibinfo
  {author} {\bibfnamefont {E.}~\bibnamefont {Solano}},\ }\bibfield  {title}
  {\bibinfo {title} {Ultrastrong coupling regimes of light-matter
  interaction},\ }\href@noop {} {\bibfield  {journal} {\bibinfo  {journal}
  {Reviews of Modern Physics}\ }\textbf {\bibinfo {volume} {91}},\ \bibinfo
  {pages} {025005} (\bibinfo {year} {2019})}\BibitemShut {NoStop}%
\bibitem [{\citenamefont {Ebbesen}(2016)}]{ebbesen16}%
  \BibitemOpen
  \bibfield  {author} {\bibinfo {author} {\bibfnamefont {T.~W.}\ \bibnamefont
  {Ebbesen}},\ }\bibfield  {title} {\bibinfo {title} {Hybrid light-matter
  states in a molecular and material science perspective},\ }\href@noop {}
  {\bibfield  {journal} {\bibinfo  {journal} {Acc. Chem. Res.}\ }\textbf
  {\bibinfo {volume} {49}},\ \bibinfo {pages} {2403} (\bibinfo {year}
  {2016})}\BibitemShut {NoStop}%
\bibitem [{\citenamefont {Garcia-Vidal}\ \emph {et~al.}(2021)\citenamefont
  {Garcia-Vidal}, \citenamefont {Ciuti},\ and\ \citenamefont
  {Ebbesen}}]{garcia21}%
  \BibitemOpen
  \bibfield  {author} {\bibinfo {author} {\bibfnamefont {F.~J.}\ \bibnamefont
  {Garcia-Vidal}}, \bibinfo {author} {\bibfnamefont {C.}~\bibnamefont
  {Ciuti}},\ and\ \bibinfo {author} {\bibfnamefont {T.~W.}\ \bibnamefont
  {Ebbesen}},\ }\bibfield  {title} {\bibinfo {title} {Manipulating matter by
  strong coupling to vacuum fields},\ }\href@noop {} {\bibfield  {journal}
  {\bibinfo  {journal} {Science}\ }\textbf {\bibinfo {volume} {373}},\ \bibinfo
  {pages} {6551} (\bibinfo {year} {2021})}\BibitemShut {NoStop}%
\bibitem [{\citenamefont {Tavis}\ and\ \citenamefont
  {Cummings}(1968)}]{tavis68}%
  \BibitemOpen
  \bibfield  {author} {\bibinfo {author} {\bibfnamefont {M.}~\bibnamefont
  {Tavis}}\ and\ \bibinfo {author} {\bibfnamefont {F.~W.}\ \bibnamefont
  {Cummings}},\ }\bibfield  {title} {\bibinfo {title} {{Exact Solution for an
  $N$-Molecule--Radiation-Field Hamiltonian}},\ }\href
  {https://doi.org/10.1103/PhysRev.170.379} {\bibfield  {journal} {\bibinfo
  {journal} {Phys. Rev.}\ }\textbf {\bibinfo {volume} {170}},\ \bibinfo {pages}
  {379} (\bibinfo {year} {1968})}\BibitemShut {NoStop}%
\bibitem [{\citenamefont {Dicke}(1954)}]{dicke54}%
  \BibitemOpen
  \bibfield  {author} {\bibinfo {author} {\bibfnamefont {R.~H.}\ \bibnamefont
  {Dicke}},\ }\bibfield  {title} {\bibinfo {title} {Coherence in spontaneous
  radiation processes},\ }\href@noop {} {\bibfield  {journal} {\bibinfo
  {journal} {Phys. Rev.}\ }\textbf {\bibinfo {volume} {93}},\ \bibinfo {pages}
  {99} (\bibinfo {year} {1954})}\BibitemShut {NoStop}%
\bibitem [{\citenamefont {Power}\ and\ \citenamefont {Zienau}(1959)}]{power59}%
  \BibitemOpen
  \bibfield  {author} {\bibinfo {author} {\bibfnamefont {E.~A.}\ \bibnamefont
  {Power}}\ and\ \bibinfo {author} {\bibfnamefont {S.}~\bibnamefont {Zienau}},\
  }\bibfield  {title} {\bibinfo {title} {Coulomb gauge in non-relativistic
  quantum electro-dynamics and the shape of spectral lines},\ }\href@noop {}
  {\bibfield  {journal} {\bibinfo  {journal} {Phil. Trans. R. Soc. Lond. A}\
  }\textbf {\bibinfo {volume} {251}},\ \bibinfo {pages} {427} (\bibinfo {year}
  {1959})}\BibitemShut {NoStop}%
\bibitem [{\citenamefont {Stegmann}\ \emph {et~al.}(2022)\citenamefont
  {Stegmann}, \citenamefont {Gupta}, \citenamefont {Haran},\ and\ \citenamefont
  {Cao}}]{cao205}%
  \BibitemOpen
  \bibfield  {author} {\bibinfo {author} {\bibfnamefont {P.}~\bibnamefont
  {Stegmann}}, \bibinfo {author} {\bibfnamefont {S.~N.}\ \bibnamefont {Gupta}},
  \bibinfo {author} {\bibfnamefont {G.}~\bibnamefont {Haran}},\ and\ \bibinfo
  {author} {\bibfnamefont {J.}~\bibnamefont {Cao}},\ }\bibfield  {title}
  {\bibinfo {title} {Higher-order photon statistics as a new tool to reveal
  hidden excited states in a plasmonic cavity},\ }\href
  {https://doi.org/10.1021/acsphotonics.2c00375} {\bibfield  {journal}
  {\bibinfo  {journal} {ACS Photonics}\ }\textbf {\bibinfo {volume} {9}},\
  \bibinfo {pages} {2119–2127} (\bibinfo {year} {2022})}\BibitemShut
  {NoStop}%
\bibitem [{\citenamefont {Tutunnikov}\ \emph
  {et~al.}(2025{\natexlab{a}})\citenamefont {Tutunnikov}, \citenamefont
  {Rokaj}, \citenamefont {Cao},\ and\ \citenamefont {Sadeghpour}}]{cao219}%
  \BibitemOpen
  \bibfield  {author} {\bibinfo {author} {\bibfnamefont {I.}~\bibnamefont
  {Tutunnikov}}, \bibinfo {author} {\bibfnamefont {V.}~\bibnamefont {Rokaj}},
  \bibinfo {author} {\bibfnamefont {J.}~\bibnamefont {Cao}},\ and\ \bibinfo
  {author} {\bibfnamefont {H.}~\bibnamefont {Sadeghpour}},\ }\bibfield  {title}
  {\bibinfo {title} {Dynamical generation and transfer of nonclassical states
  in strongly interacting light-matter systems in cavities},\ }\href@noop {}
  {\bibfield  {journal} {\bibinfo  {journal} {Quantum Science and Technology}\
  }\textbf {\bibinfo {volume} {10}},\ \bibinfo {pages} {025002} (\bibinfo
  {year} {2025}{\natexlab{a}})}\BibitemShut {NoStop}%
\bibitem [{\citenamefont {Jaynes}\ and\ \citenamefont
  {Cummings}(1963)}]{jaynes63}%
  \BibitemOpen
  \bibfield  {author} {\bibinfo {author} {\bibfnamefont {E.~T.}\ \bibnamefont
  {Jaynes}}\ and\ \bibinfo {author} {\bibfnamefont {F.~W.}\ \bibnamefont
  {Cummings}},\ }\bibfield  {title} {\bibinfo {title} {Comparison of quantum
  and semiclassical radiation theory with application to the beam maser},\
  }\href {https://doi.org/10.1109/PROC.1963.1664} {\bibfield  {journal}
  {\bibinfo  {journal} {Proceedings of the IEEE}\ }\textbf {\bibinfo {volume}
  {51}},\ \bibinfo {pages} {89} (\bibinfo {year} {1963})}\BibitemShut {NoStop}%
\bibitem [{\citenamefont {Gross}\ and\ \citenamefont
  {Haroche}(1982)}]{gross82}%
  \BibitemOpen
  \bibfield  {author} {\bibinfo {author} {\bibfnamefont {M.}~\bibnamefont
  {Gross}}\ and\ \bibinfo {author} {\bibfnamefont {S.}~\bibnamefont
  {Haroche}},\ }\bibfield  {title} {\bibinfo {title} {Superradiance: an essay
  on the theory of collective spontaneous emission},\ }\href@noop {} {\bibfield
   {journal} {\bibinfo  {journal} {Physics Reports}\ }\textbf {\bibinfo
  {volume} {93}},\ \bibinfo {pages} {301} (\bibinfo {year} {1982})}\BibitemShut
  {NoStop}%
\bibitem [{\citenamefont {Shore}\ and\ \citenamefont {Knight}(1993)}]{shore93}%
  \BibitemOpen
  \bibfield  {author} {\bibinfo {author} {\bibfnamefont {B.~W.}\ \bibnamefont
  {Shore}}\ and\ \bibinfo {author} {\bibfnamefont {P.~L.}\ \bibnamefont
  {Knight}},\ }\bibfield  {title} {\bibinfo {title} {Topical review: the
  jaynes--cummings model},\ }\href {https://doi.org/10.1080/09500349314551321}
  {\bibfield  {journal} {\bibinfo  {journal} {Journal of Modern Optics}\
  }\textbf {\bibinfo {volume} {40}},\ \bibinfo {pages} {1195} (\bibinfo {year}
  {1993})}\BibitemShut {NoStop}%
\bibitem [{\citenamefont {Emary}\ and\ \citenamefont
  {Brandes}(2003)}]{emary03}%
  \BibitemOpen
  \bibfield  {author} {\bibinfo {author} {\bibfnamefont {C.}~\bibnamefont
  {Emary}}\ and\ \bibinfo {author} {\bibfnamefont {T.}~\bibnamefont
  {Brandes}},\ }\bibfield  {title} {\bibinfo {title} {Chaos and the quantum
  phase transition in the dicke model},\ }\href@noop {} {\bibfield  {journal}
  {\bibinfo  {journal} {Phys. Rev. E}\ }\textbf {\bibinfo {volume} {67}},\
  \bibinfo {pages} {066203} (\bibinfo {year} {2003})}\BibitemShut {NoStop}%
\bibitem [{\citenamefont {Schafer}\ \emph {et~al.}(2020)\citenamefont
  {Schafer}, \citenamefont {Ruggenthaler}, \citenamefont {Rokaj},\ and\
  \citenamefont {Rubio}}]{schafer20a}%
  \BibitemOpen
  \bibfield  {author} {\bibinfo {author} {\bibfnamefont {C.}~\bibnamefont
  {Schafer}}, \bibinfo {author} {\bibfnamefont {M.}~\bibnamefont
  {Ruggenthaler}}, \bibinfo {author} {\bibfnamefont {V.}~\bibnamefont
  {Rokaj}},\ and\ \bibinfo {author} {\bibfnamefont {A.}~\bibnamefont {Rubio}},\
  }\bibfield  {title} {\bibinfo {title} {Relevance of the quadratic diamagnetic
  and self-polarization terms in cavity quantum electrodynamics},\ }\href@noop
  {} {\bibfield  {journal} {\bibinfo  {journal} {ACS Photonics}\ }\textbf
  {\bibinfo {volume} {7}},\ \bibinfo {pages} {975} (\bibinfo {year}
  {2020})}\BibitemShut {NoStop}%
\bibitem [{\citenamefont {Taylor}\ \emph {et~al.}(2020)\citenamefont {Taylor},
  \citenamefont {A~Mandal}, \citenamefont {W},\ and\ \citenamefont
  {Huo}}]{taylor20}%
  \BibitemOpen
  \bibfield  {author} {\bibinfo {author} {\bibfnamefont {M.}~\bibnamefont
  {Taylor}}, \bibinfo {author} {\bibfnamefont {A.}~\bibnamefont {A~Mandal}},
  \bibinfo {author} {\bibfnamefont {Z.}~\bibnamefont {W}},\ and\ \bibinfo
  {author} {\bibfnamefont {P.}~\bibnamefont {Huo}},\ }\bibfield  {title}
  {\bibinfo {title} {Resolution of gauge ambiguities in molecular cavity
  quantum electrodynamics},\ }\href@noop {} {\bibfield  {journal} {\bibinfo
  {journal} {Phys. Rev. Lett.}\ }\textbf {\bibinfo {volume} {125}},\ \bibinfo
  {pages} {123602} (\bibinfo {year} {2020})}\BibitemShut {NoStop}%
\bibitem [{\citenamefont {Cao}\ and\ \citenamefont {Pollak}(2025)}]{cao221}%
  \BibitemOpen
  \bibfield  {author} {\bibinfo {author} {\bibfnamefont {J.}~\bibnamefont
  {Cao}}\ and\ \bibinfo {author} {\bibfnamefont {E.}~\bibnamefont {Pollak}},\
  }\bibfield  {title} {\bibinfo {title} {Cavity-induced quantum interference
  and collective interactions in van der waals systems},\ }\href@noop {}
  {\bibfield  {journal} {\bibinfo  {journal} {J. Phys. Chem. Lett.}\ }\textbf
  {\bibinfo {volume} {16}},\ \bibinfo {pages} {5466} (\bibinfo {year}
  {2025})}\BibitemShut {NoStop}%
\bibitem [{\citenamefont {Majer}\ \emph {et~al.}(2007)\citenamefont {Majer},
  \citenamefont {Chow}, \citenamefont {Gambetta}, \citenamefont {Koch},
  \citenamefont {Johnson}, \citenamefont {Schreier}, \citenamefont {Frunzio},
  \citenamefont {Schuster}, \citenamefont {Houck}, \citenamefont {Wallraff},
  \citenamefont {Blais}, \citenamefont {Devoret}, \citenamefont {Girvin},\ and\
  \citenamefont {Schoelkopf}}]{majer07}%
  \BibitemOpen
  \bibfield  {author} {\bibinfo {author} {\bibfnamefont {J.}~\bibnamefont
  {Majer}}, \bibinfo {author} {\bibfnamefont {J.~M.}\ \bibnamefont {Chow}},
  \bibinfo {author} {\bibfnamefont {J.~M.}\ \bibnamefont {Gambetta}}, \bibinfo
  {author} {\bibfnamefont {J.}~\bibnamefont {Koch}}, \bibinfo {author}
  {\bibfnamefont {B.~R.}\ \bibnamefont {Johnson}}, \bibinfo {author}
  {\bibfnamefont {J.~A.}\ \bibnamefont {Schreier}}, \bibinfo {author}
  {\bibfnamefont {L.}~\bibnamefont {Frunzio}}, \bibinfo {author} {\bibfnamefont
  {D.~I.}\ \bibnamefont {Schuster}}, \bibinfo {author} {\bibfnamefont {A.~A.}\
  \bibnamefont {Houck}}, \bibinfo {author} {\bibfnamefont {A.}~\bibnamefont
  {Wallraff}}, \bibinfo {author} {\bibfnamefont {A.}~\bibnamefont {Blais}},
  \bibinfo {author} {\bibfnamefont {M.~H.}\ \bibnamefont {Devoret}}, \bibinfo
  {author} {\bibfnamefont {S.~M.}\ \bibnamefont {Girvin}},\ and\ \bibinfo
  {author} {\bibfnamefont {R.~J.}\ \bibnamefont {Schoelkopf}},\ }\bibfield
  {title} {\bibinfo {title} {Coupling superconducting qubits via a cavity
  bus},\ }\href@noop {} {\bibfield  {journal} {\bibinfo  {journal} {Nature}\
  }\textbf {\bibinfo {volume} {449}},\ \bibinfo {pages} {443} (\bibinfo {year}
  {2007})}\BibitemShut {NoStop}%
\bibitem [{\citenamefont {Place}\ \emph {et~al.}(2021)\citenamefont {Place},
  \citenamefont {Rodgers}, \citenamefont {Mundada}, \citenamefont {Smitham},
  \citenamefont {Fitzpatrick}, \citenamefont {Leng}, \citenamefont {Premkumar},
  \citenamefont {Bryon}, \citenamefont {Vrajitoarea}, \citenamefont {Sussman},
  \citenamefont {Cheng}, \citenamefont {Madhavan}, \citenamefont {Babla},
  \citenamefont {Le}, \citenamefont {Gang}, \citenamefont {Jäck},
  \citenamefont {Gyenis}, \citenamefont {Yao}, \citenamefont {Cava},
  \citenamefont {de~Leon},\ and\ \citenamefont {Houck}}]{place21}%
  \BibitemOpen
  \bibfield  {author} {\bibinfo {author} {\bibfnamefont {A.~P.~M.}\
  \bibnamefont {Place}}, \bibinfo {author} {\bibfnamefont {L.~V.}\ \bibnamefont
  {Rodgers}}, \bibinfo {author} {\bibfnamefont {P.}~\bibnamefont {Mundada}},
  \bibinfo {author} {\bibfnamefont {B.~M.}\ \bibnamefont {Smitham}}, \bibinfo
  {author} {\bibfnamefont {M.}~\bibnamefont {Fitzpatrick}}, \bibinfo {author}
  {\bibfnamefont {Z.}~\bibnamefont {Leng}}, \bibinfo {author} {\bibfnamefont
  {A.}~\bibnamefont {Premkumar}}, \bibinfo {author} {\bibfnamefont
  {J.}~\bibnamefont {Bryon}}, \bibinfo {author} {\bibfnamefont
  {A.}~\bibnamefont {Vrajitoarea}}, \bibinfo {author} {\bibfnamefont
  {S.}~\bibnamefont {Sussman}}, \bibinfo {author} {\bibfnamefont
  {G.}~\bibnamefont {Cheng}}, \bibinfo {author} {\bibfnamefont
  {T.}~\bibnamefont {Madhavan}}, \bibinfo {author} {\bibfnamefont {H.~K.}\
  \bibnamefont {Babla}}, \bibinfo {author} {\bibfnamefont {X.~H.}\ \bibnamefont
  {Le}}, \bibinfo {author} {\bibfnamefont {Y.}~\bibnamefont {Gang}}, \bibinfo
  {author} {\bibfnamefont {B.}~\bibnamefont {Jäck}}, \bibinfo {author}
  {\bibfnamefont {A.}~\bibnamefont {Gyenis}}, \bibinfo {author} {\bibfnamefont
  {N.}~\bibnamefont {Yao}}, \bibinfo {author} {\bibfnamefont {R.~J.}\
  \bibnamefont {Cava}}, \bibinfo {author} {\bibfnamefont {N.~P.}\ \bibnamefont
  {de~Leon}},\ and\ \bibinfo {author} {\bibfnamefont {A.~A.}\ \bibnamefont
  {Houck}},\ }\bibfield  {title} {\bibinfo {title} {New material platform for
  superconducting transmon qubits with coherence times exceeding 0.3
  milliseconds},\ }\href@noop {} {\bibfield  {journal} {\bibinfo  {journal}
  {Nature Communications}\ }\textbf {\bibinfo {volume} {12}},\ \bibinfo {pages}
  {1779} (\bibinfo {year} {2021})}\BibitemShut {NoStop}%
\bibitem [{\citenamefont {George}\ \emph {et~al.}(2016)\citenamefont {George},
  \citenamefont {Chervy}, \citenamefont {Shalabney}, \citenamefont {Devaux},
  \citenamefont {Hiura}, \citenamefont {Genet},\ and\ \citenamefont
  {Ebbesen}}]{george16}%
  \BibitemOpen
  \bibfield  {author} {\bibinfo {author} {\bibfnamefont {J.}~\bibnamefont
  {George}}, \bibinfo {author} {\bibfnamefont {T.}~\bibnamefont {Chervy}},
  \bibinfo {author} {\bibfnamefont {A.}~\bibnamefont {Shalabney}}, \bibinfo
  {author} {\bibfnamefont {E.}~\bibnamefont {Devaux}}, \bibinfo {author}
  {\bibfnamefont {H.}~\bibnamefont {Hiura}}, \bibinfo {author} {\bibfnamefont
  {C.}~\bibnamefont {Genet}},\ and\ \bibinfo {author} {\bibfnamefont
  {T.}~\bibnamefont {Ebbesen}},\ }\bibfield  {title} {\bibinfo {title}
  {Multiple rabi splittings under ultra-strong vibrational coupling},\
  }\href@noop {} {\bibfield  {journal} {\bibinfo  {journal} {Phys. Rev. Lett.}\
  }\textbf {\bibinfo {volume} {117}},\ \bibinfo {pages} {153601} (\bibinfo
  {year} {2016})}\BibitemShut {NoStop}%
\bibitem [{\citenamefont {Xiang}\ \emph {et~al.}(2018)\citenamefont {Xiang},
  \citenamefont {Ribeiro}, \citenamefont {Dunkelberger}, \citenamefont
  {J.Wang}, \citenamefont {Li}, \citenamefont {Simpkins}, \citenamefont
  {Owrutsky}, \citenamefont {Yuen-Zhou},\ and\ \citenamefont
  {Xiong}}]{xiang18}%
  \BibitemOpen
  \bibfield  {author} {\bibinfo {author} {\bibfnamefont {B.}~\bibnamefont
  {Xiang}}, \bibinfo {author} {\bibfnamefont {R.~F.}\ \bibnamefont {Ribeiro}},
  \bibinfo {author} {\bibfnamefont {A.~D.}\ \bibnamefont {Dunkelberger}},
  \bibinfo {author} {\bibnamefont {J.Wang}}, \bibinfo {author} {\bibfnamefont
  {Y.}~\bibnamefont {Li}}, \bibinfo {author} {\bibfnamefont {B.~S.}\
  \bibnamefont {Simpkins}}, \bibinfo {author} {\bibfnamefont {J.~C.}\
  \bibnamefont {Owrutsky}}, \bibinfo {author} {\bibfnamefont {J.}~\bibnamefont
  {Yuen-Zhou}},\ and\ \bibinfo {author} {\bibfnamefont {W.}~\bibnamefont
  {Xiong}},\ }\bibfield  {title} {\bibinfo {title} {Two-dimensional infrared
  spectroscopy of vibrational polaritons},\ }\href@noop {} {\bibfield
  {journal} {\bibinfo  {journal} {Proc. Natl. Acad. Sci. U.S.A.}\ }\textbf
  {\bibinfo {volume} {115}},\ \bibinfo {pages} {4845} (\bibinfo {year}
  {2018})}\BibitemShut {NoStop}%
\bibitem [{\citenamefont {Engelhardt}\ and\ \citenamefont
  {Cao}(2022)}]{cao204}%
  \BibitemOpen
  \bibfield  {author} {\bibinfo {author} {\bibfnamefont {G.}~\bibnamefont
  {Engelhardt}}\ and\ \bibinfo {author} {\bibfnamefont {J.}~\bibnamefont
  {Cao}},\ }\bibfield  {title} {\bibinfo {title} {Unusual dynamical properties
  of disordered polaritons in microcavities},\ }\href
  {http://dx.doi.org/10.1103/PhysRevB.105.064205} {\bibfield  {journal}
  {\bibinfo  {journal} {Phys. Rev. B}\ }\textbf {\bibinfo {volume} {105}},\
  \bibinfo {pages} {064205/1} (\bibinfo {year} {2022})}\BibitemShut {NoStop}%
\bibitem [{\citenamefont {Engelhardt}\ and\ \citenamefont
  {Cao}(2023)}]{cao211}%
  \BibitemOpen
  \bibfield  {author} {\bibinfo {author} {\bibfnamefont {G.}~\bibnamefont
  {Engelhardt}}\ and\ \bibinfo {author} {\bibfnamefont {J.}~\bibnamefont
  {Cao}},\ }\bibfield  {title} {\bibinfo {title} {Polarition localization and
  spectroscopic properties of disordered quantum emitters in spatially-extended
  microcavities},\ }\href@noop {} {\bibfield  {journal} {\bibinfo  {journal}
  {Phys. Rev. Letts.}\ }\textbf {\bibinfo {volume} {130}},\ \bibinfo {pages}
  {213602} (\bibinfo {year} {2023})}\BibitemShut {NoStop}%
\bibitem [{\citenamefont {Tutunnikov}\ \emph
  {et~al.}(2025{\natexlab{b}})\citenamefont {Tutunnikov}, \citenamefont
  {Qutubuddin}, \citenamefont {Sadeghpour},\ and\ \citenamefont
  {Cao}}]{cao223}%
  \BibitemOpen
  \bibfield  {author} {\bibinfo {author} {\bibfnamefont {I.}~\bibnamefont
  {Tutunnikov}}, \bibinfo {author} {\bibfnamefont {M.}~\bibnamefont
  {Qutubuddin}}, \bibinfo {author} {\bibfnamefont {H.}~\bibnamefont
  {Sadeghpour}},\ and\ \bibinfo {author} {\bibfnamefont {J.}~\bibnamefont
  {Cao}},\ }\bibfield  {title} {\bibinfo {title} {Characterization of polariton
  dynamics in a multimode cavity: Noise-enhanced ballistic expansion},\
  }\href@noop {} {\bibfield  {journal} {\bibinfo  {journal} {arXiv preprint
  arXiv:2410.11051}\ } (\bibinfo {year} {2025}{\natexlab{b}})}\BibitemShut
  {NoStop}%
\bibitem [{\citenamefont {Wu}\ \emph {et~al.}(2024)\citenamefont {Wu},
  \citenamefont {Cerrillo},\ and\ \citenamefont {Cao}}]{cao217}%
  \BibitemOpen
  \bibfield  {author} {\bibinfo {author} {\bibfnamefont {A.}~\bibnamefont
  {Wu}}, \bibinfo {author} {\bibfnamefont {J.}~\bibnamefont {Cerrillo}},\ and\
  \bibinfo {author} {\bibfnamefont {J.}~\bibnamefont {Cao}},\ }\bibfield
  {title} {\bibinfo {title} {Extracting kinetic information from short-time
  trajectories: relaxation and disorder of lossy cavity polaritons},\
  }\href@noop {} {\bibfield  {journal} {\bibinfo  {journal} {Nanophotonics}\
  }\textbf {\bibinfo {volume} {13}},\ \bibinfo {pages} {2575} (\bibinfo {year}
  {2024})}\BibitemShut {NoStop}%
\bibitem [{\citenamefont {Hoffmann}\ \emph {et~al.}(2025)\citenamefont
  {Hoffmann}, \citenamefont {Schäfer}, \citenamefont {Rubio}, \citenamefont
  {Kelly},\ and\ \citenamefont {Appel}}]{hoffmann19b}%
  \BibitemOpen
  \bibfield  {author} {\bibinfo {author} {\bibfnamefont {N.}~\bibnamefont
  {Hoffmann}}, \bibinfo {author} {\bibfnamefont {C.}~\bibnamefont {Schäfer}},
  \bibinfo {author} {\bibfnamefont {A.}~\bibnamefont {Rubio}}, \bibinfo
  {author} {\bibfnamefont {A.}~\bibnamefont {Kelly}},\ and\ \bibinfo {author}
  {\bibfnamefont {H.}~\bibnamefont {Appel}},\ }\bibfield  {title} {\bibinfo
  {title} {Capturing vacuum fluctuations and photon correlations in cavity
  quantum electrodynamics with multitrajectory ehrenfest dynamics},\
  }\href@noop {} {\bibfield  {journal} {\bibinfo  {journal} {Physical Review
  A}\ }\textbf {\bibinfo {volume} {99}},\ \bibinfo {pages} {063819} (\bibinfo
  {year} {2025})}\BibitemShut {NoStop}%
\bibitem [{\citenamefont {Li}\ \emph {et~al.}(2019)\citenamefont {Li},
  \citenamefont {H.T.}, \citenamefont {Nitzan},\ and\ \citenamefont
  {Subotnik}}]{li19}%
  \BibitemOpen
  \bibfield  {author} {\bibinfo {author} {\bibfnamefont {T.~E.}\ \bibnamefont
  {Li}}, \bibinfo {author} {\bibfnamefont {C.}~\bibnamefont {H.T.}}, \bibinfo
  {author} {\bibfnamefont {A.}~\bibnamefont {Nitzan}},\ and\ \bibinfo {author}
  {\bibfnamefont {J.~E.}\ \bibnamefont {Subotnik}},\ }\bibfield  {title}
  {\bibinfo {title} {Understanding the nature of mean-field semiclassical
  light-matter dynamics: An investigation of energy transfer, electron-electron
  correlations, external driving, and long-time detailed balance},\ }\href@noop
  {} {\bibfield  {journal} {\bibinfo  {journal} {Physical Review A}\ }\textbf
  {\bibinfo {volume} {100}},\ \bibinfo {pages} {062509} (\bibinfo {year}
  {2019})}\BibitemShut {NoStop}%
\bibitem [{\citenamefont {Hsieh}\ and\ \citenamefont
  {Tempelaar}(2025)}]{hsieh25}%
  \BibitemOpen
  \bibfield  {author} {\bibinfo {author} {\bibfnamefont {M.-H.}\ \bibnamefont
  {Hsieh}}\ and\ \bibinfo {author} {\bibfnamefont {R.}~\bibnamefont
  {Tempelaar}},\ }\bibfield  {title} {\bibinfo {title} {Mixed
  quantum–classical dynamics yields anharmonic rabi oscillations},\
  }\href@noop {} {\bibfield  {journal} {\bibinfo  {journal} {Journal of
  Chemical Physics}\ }\textbf {\bibinfo {volume} {162}},\ \bibinfo {pages}
  {224109} (\bibinfo {year} {2025})}\BibitemShut {NoStop}%
\bibitem [{\citenamefont {Wu}\ and\ \citenamefont {Cao}(2001)}]{cao54}%
  \BibitemOpen
  \bibfield  {author} {\bibinfo {author} {\bibfnamefont {J.}~\bibnamefont
  {Wu}}\ and\ \bibinfo {author} {\bibfnamefont {J.}~\bibnamefont {Cao}},\
  }\bibfield  {title} {\bibinfo {title} {Linear and nonlinear response
  functions of the morse oscillator: Classical divergence and the uncertainty
  principle},\ }\href@noop {} {\bibfield  {journal} {\bibinfo  {journal} {J.
  Chem. Phys.}\ }\textbf {\bibinfo {volume} {115}},\ \bibinfo {pages} {5381}
  (\bibinfo {year} {2001})}\BibitemShut {NoStop}%
\bibitem [{\citenamefont {Kryvohuz}\ and\ \citenamefont {Cao}(2005)}]{cao86}%
  \BibitemOpen
  \bibfield  {author} {\bibinfo {author} {\bibfnamefont {M.}~\bibnamefont
  {Kryvohuz}}\ and\ \bibinfo {author} {\bibfnamefont {J.}~\bibnamefont {Cao}},\
  }\bibfield  {title} {\bibinfo {title} {Quantum-classical correspondence in
  response theory},\ }\href@noop {} {\bibfield  {journal} {\bibinfo  {journal}
  {Phys. Rev. Lett.}\ }\textbf {\bibinfo {volume} {95}},\ \bibinfo {pages}
  {180405 1} (\bibinfo {year} {2005})}\BibitemShut {NoStop}%
\bibitem [{\citenamefont {Gruenbaum}\ and\ \citenamefont
  {Loring}(2008)}]{gruenbaum08}%
  \BibitemOpen
  \bibfield  {author} {\bibinfo {author} {\bibfnamefont {S.~M.}\ \bibnamefont
  {Gruenbaum}}\ and\ \bibinfo {author} {\bibfnamefont {R.~F.}\ \bibnamefont
  {Loring}},\ }\bibfield  {title} {\bibinfo {title} {Interference and
  quantization in semiclassical response functions},\ }\href@noop {} {\bibfield
   {journal} {\bibinfo  {journal} {Journal of Chemical Physics}\ }\textbf
  {\bibinfo {volume} {128}},\ \bibinfo {pages} {124106} (\bibinfo {year}
  {2008})}\BibitemShut {NoStop}%
\bibitem [{\citenamefont {Dutta}\ and\ \citenamefont
  {Reppert}(2025)}]{dutta25}%
  \BibitemOpen
  \bibfield  {author} {\bibinfo {author} {\bibfnamefont {R.}~\bibnamefont
  {Dutta}}\ and\ \bibinfo {author} {\bibfnamefont {M.}~\bibnamefont
  {Reppert}},\ }\bibfield  {title} {\bibinfo {title} {Quantum and classical
  effects in system-bath correlations and optical line shapes},\ }\href@noop {}
  {\bibfield  {journal} {\bibinfo  {journal} {Physical Review A}\ }\textbf
  {\bibinfo {volume} {11}},\ \bibinfo {pages} {022210} (\bibinfo {year}
  {2025})}\BibitemShut {NoStop}%
\end{thebibliography}%

\end{document}